\documentclass[twocolumn,showpacs,prd]{revtex4}
\usepackage{graphicx}

\begin{document}
\title{$Z'$ signal from the LEP2 data}

\author{A.V. Gulov}
 \email{gulov@ff.dsu.dp.ua}
\author{V.V. Skalozub}
 \email{skalozub@ff.dsu.dp.ua}
\affiliation{
 Dnepropetrovsk National University, Dnepropetrovsk 49050, Ukraine
}
\date{\today}
\begin{abstract}
The many-parametric fit of the LEP2 data on $e^+e^-\to e^+e^-$,
$\mu^+\mu^-$, $\tau^+\tau^-$ processes is performed to estimate
signals of the Abelian $Z'$-boson beyond the standard model. The
model-independent relations between the $Z'$ couplings to the
standard model particles allow to describe the $Z'$ effects in
lepton processes by 4 independent parameters. No signal is found
by the complete LEP2 data set, and the 1.3$\sigma$ signal is
detected by the fit of the backward bins. The $Z'$ couplings to
the vector and axial-vector lepton currents are constrained. The
comparisons with the one-parameter fits and with the LEP1
experiments are performed.
\end{abstract}

\pacs{14.70.Pw, 13.66.-a}

\maketitle

\section{Introduction}

The precision test of the standard model (SM) at LEP gave a
possibility not only to determine all the parameters and particle
masses at the level of radiative corrections but also afforded an
opportunity for searching for signals of new heavy particles
beyond the energy scale of it. On the base of the LEP2 experiments
the low bounds on parameters of various models extending the SM
have been estimated and the scale of new physics was obtained
\cite{EWWG,OPAL,DELPHI}. Although no new particles were
discovered, a general believe is that the energy scale of new
physics to be of order 1 TeV, that may serve as a guide for
experiments at the LHC. In this situation, any information about
new heavy particles obtained on the base of the present day data
is desirable and important.

A lot of extended models includes a so-called $Z'$ gauge boson --
a massive neutral vector particle associated with an extra $U(1)$
subgroup of the underlying group. Searching for this particle in
either model-dependent or model-independent approaches is widely
discussed in the literature \cite{Leike}. In the papers of the
present authors \cite{EPJC2000,YAF2004,PRD2004} the new method and
the observables for the model-independent search for the
$Z'$-boson were proposed and applied to analyze the LEP2
experiment data. In contrast to other model-independent searches,
our approach gives a possibility to pick out uniquely this virtual
state and determine its characteristics. It is based on two
ingredients: 1) The relations between the low-energy parameters
are derived from the renormalization group (RG) equation for the
fermion scattering amplitude and therefore called the RG
relations. Due to them the number of unknown $Z'$ parameters to be
measured considerably decreases. 2) The kinematic properties of
the $Z'$ signals in various scattering process exhibit themselves
when these relations are accounted for. In these papers the
one-parametric observables were introduced and the signals of the
$Z'$ have been determined at the 1$\sigma$ confidence level (CL)
in the $e^+e^-\to\mu^+\mu^-$ process, and at the 2$\sigma$ CL in
the Bhabha process. The $Z'$ mass was estimated to be 1--1.2 TeV
that gives a good chance to discover the particle at the LHC.

Recently the final data of the LEP collaborations DELPHI and OPAL
\cite{OPAL,DELPHI} were published and new more precise estimates
could be obtained. In the present paper we update the results of
the one-parameter fit and perform the complete many-parametric fit
of the LEP2 data to estimate a possible signal of the $Z'$-boson.
Usually, in a many-parametric fit the uncertainty of the result
incerases drastically because of extra parameters. On the
contrary, in our approach due to the RG relations between the
low-energy couplings there are only 2-3 independent parameters for
the LEP scattering processes. Therefore, we believe that an
inevitable increase of confidence areas (CA) in the
many-parametric space could be compensated by accounting for all
the accessible experimental information. As it will be shown, the
uncertainty of the many-parametric fit can be comparable with the
uncertainty of the previous one-parametric fits in Refs.
\cite{YAF2004,PRD2004}. In this approach the combined data fit for
all lepton processes is also possible.

\section{The Abelian $Z'$ boson at low energies}

Let us adduce a necessary information about the Abelian
$Z'$-boson. This particle is predicted by a number of grand
unification models. Among them the $E_6$ and $SO(10)$ based models
\cite{Hewett} (for instance, LR, $\chi-\psi$ and so on) are often
discussed in the literature. In all the models, the Abelian
$Z'$-boson is described by a low-energy $\tilde{U}(1)$ gauge
subgroup originated in some symmetry breaking pattern.

At low energies, the $Z'$-boson can manifest itself by means of
the couplings to the SM fermions and scalars as a virtual
intermediate state. Moreover, the $Z$-boson couplings are also
modified due to a $Z$--$Z'$ mixing. In principle, arbitrary
effective $Z'$ interactions to the SM fields could be considered
at low energies. However, the couplings of non-renormalizable
types have to be suppressed by heavy mass scales because of
decoupling. Therefore, significant signals beyond the SM can be
inspired by the couplings of renormalizable types. Such couplings
can be derived by adding the new $\tilde{U}(1)$-terms to the
electroweak covariant derivatives $D^\mathrm{ew}$ in the
Lagrangian \cite{Leike}
\begin{eqnarray}\label{1a}
 {\cal L}&=& \left|\left( D^\mathrm{ew}_\mu -
  i\frac{\tilde{y}_\phi}{2}\tilde{Z}_\mu \right)\phi\right|^2 +
  \nonumber\\&&+
   i\sum\limits_{f=f_L,f_R}\bar{f}{\gamma^\mu}
  \left(
  D^\mathrm{ew}_\mu -
  i\frac{\tilde{y}_f}{2}\tilde{Z}_\mu
  \right)f,
\end{eqnarray}
where $\phi$ is the SM scalar doublet; $f_L$, $f_R$ are the SM
left-handed fermion doublets and right-handed fermion singlets;
$\tilde{Z}_\mu$ denotes the $\tilde{U}(1)$ symmetry eigenstate;
and $\tilde{y}_\phi$, $\tilde{y}_{f_L}$ and $\tilde{y}_{f_R}$ mean
the unknown couplings characterizing the model beyond the SM.
Instead of the couplings to the left-handed and right-handed
fermion states it is convenient to introduce the couplings to the
axial-vector and vector currents:
$a_f=(\tilde{y}_{f_R}-\tilde{y}_{f_L})/2$,
$v_f=(\tilde{y}_{f_L}+\tilde{y}_{f_R})/2$.

The spontaneous breaking of the electroweak symmetry leads to the
$Z$--$Z'$ mixing. In case of the Abelian $Z'$-boson, the $Z$--$Z'$ mixing angle $\theta_0$
is determined by the coupling
$\tilde{y}_\phi$ as follows \cite{EPJC2000}
\begin{equation}\label{2}
\theta_0 = \frac{\sin\theta_W\cos\theta_W}{\sqrt{4\pi\alpha_\mathrm{em}}}
\frac{m^2_Z}{m^2_{Z'}} \tilde{y}_\phi
+O\left(\frac{m^4_Z}{m^4_{Z'}}\right),
\end{equation}
where $\theta_W$ is the SM Weinberg angle, and
$\alpha_\mathrm{em}$ is the electromagnetic fine structure
constant. Although the mixing angle is a small quantity of order
$m^{-2}_{Z'}$, it contributes to the $Z$-boson exchange amplitude
and cannot be neglected at the LEP energies.

The Lagrangian (\ref{1a}) leads to the following interactions between
the fermions and the $Z$ and $Z'$ mass eigenstates:
\begin{eqnarray}
{\cal L}_{Z\bar{f}f}&=&\frac{1}{2}iZ_\mu\bar{f}\gamma^\mu\left[
(v^\mathrm{SM}_{fZ}+\gamma^5 a^\mathrm{SM}_{fZ})\cos\theta_0
+\right.\nonumber\\&&\quad\left.
+(v_f+\gamma^5 a_f)\sin\theta_0 \right]f, \nonumber\\
{\cal L}_{Z'\bar{f}f}&=&\frac{1}{2}iZ'_\mu\bar{f}\gamma^\mu\left[
(v_f+\gamma^5 a_f)\cos\theta_0 -\right.\nonumber\\&&\quad\left.
-(v^\mathrm{SM}_{fZ}+\gamma^5
a^\mathrm{SM}_{fZ})\sin\theta_0\right]f,
\end{eqnarray}
where $f$ is an arbitrary SM fermion state; $v^\mathrm{SM}_{fZ}$,
$a^\mathrm{SM}_{fZ}$ are the SM couplings of the $Z$-boson.

In a particular model the couplings $v_f$ and $a_f$ take some
specific values. In case when the model is unknown, these
parameters and the mixing angle remain potentially arbitrary
numbers. However, this is not the case if one assumes that the
underlying extended model is a renormalizable one. As it was shown
in Ref. \cite{EPJC2000}, some of them have to be correlated due to
renormalizability. The corresponding relations are
\begin{equation}\label{3}
v_f - a_f = v_{f^\star} - a_{f^\star},\quad a_f =
T_{3,f}\tilde{y}_\phi,
\end{equation}
where $f^\star$ is the SU(2) partner of a fermion $f$, and
$T_{3,f}$ is the third component of the fermion isospin. They are
motivated by the renormalization group equations at the $Z'$
decoupling energies and also connected with the $\tilde{U}(1)$
gauge symmetry of the Lagrangian. These relations cover all the
popular models of the Abelian $Z'$ boson allowing the
model-independent searches for this particle.

The relations (\ref{3}) incorporate the most common features of
the Abelian $Z'$-boson. As it is seen, the axial-vector coupling
is universal for all the fermion flavors. So, in what follows we
will use the shorthand notation $a=a_e=a_\mu=a_\tau$. The
axial-vector coupling determines also the coupling to the scalar
doublet and, consequently, the mixing angle. As a result, the
number of independent couplings is significantly reduced.
Considering the leptonic processes $e^+e^-\to\ell^+\ell^-$
($\ell=e,\mu,\tau$), one has to keep 4 unknown couplings: $a$,
$v_e$, $v_\mu$, and $v_\tau$. Moreover, the RG relations serve to
uniquely specify a kinematic domain of deviations from the SM
predictions due to the virtual $Z'$ boson. Thereof a single
definition of the $Z'$ signal can be done.

In our analysis, as the SM values of the cross-sections we use the
quantities calculated by the LEP2 collaborations
\cite{ALEPH,DELPHI,L3,OPAL}. They account for either the one-loop
radiative corrections or initial and final state radiation effects
(together with the event selection rules, which are specific for
each experiment). The deviation from the SM is computed in the
improved Born approximation. This accuracy is sufficient for our
analysis, leading to the systematic error of fit results less than
5-10\%.

The differential cross-section of the process
$e^+e^-\to\ell^+\ell^-$ deviates from the SM value by various
quadratic combinations of couplings $a$, $v_e$, $v_\mu$, $v_\tau$.
For the Bhabha process it reads
\begin{equation}\label{4}
\frac{d\sigma}{dz}-\frac{d\sigma^\mathrm{SM}}{dz} =
f^{ee}_1(z)\frac{a^2}{m_{Z'}^2} +
f^{ee}_2(z)\frac{v_e^2}{m_{Z'}^2} +
f^{ee}_3(z)\frac{a v_e}{m_{Z'}^2},
\end{equation}
where the factors are known functions of the center-of-mass energy
and the cosine of the electron scattering angle $z$. The deviation
of the cross-section of $e^+e^-\to\mu^+\mu^-$ ($\tau^+\tau^-$)
processes has the similar form
\begin{eqnarray}\label{5}
\frac{d\sigma}{dz}-\frac{d\sigma^\mathrm{SM}}{dz} &=&
f^{\mu\mu}_1(z)\frac{a^2}{m_{Z'}^2} +
f^{\mu\mu}_2(z)\frac{v_e v_\mu}{m_{Z'}^2} +
\nonumber\\ &&+
f^{\mu\mu}_3(z)\frac{a v_e}{m_{Z'}^2} +
f^{\mu\mu}_4(z)\frac{a v_\mu}{m_{Z'}^2}.
\end{eqnarray}
This is our definition of the $Z'$ signal.

Since the $Z'$ couplings enter the cross-section together with the
inverse $Z'$ mass, it is convenient to introduce the dimensionless
couplings
\begin{equation}\label{6}
\bar{a}_f=\frac{m_Z}{\sqrt{4\pi}m_{Z'}}a_f,\quad
\bar{v}_f=\frac{m_Z}{\sqrt{4\pi}m_{Z'}}v_f,
\end{equation}
which can be constrained by experiments.

Note again, that the cross-sections in Eqs. (\ref{4})--(\ref{5})
account for the relations (\ref{3}) through the functions
$f_1(z)$, $f_3(z)$, $f_4(z)$, since the coupling $\tilde{y}_\phi$
(the mixing angle $\theta_0$) is substituted by the axial coupling
constant $a$. Usually, this dependence on the scalar field
coupling is neglected at all, when a four-fermion effective
Lagrangian is applied to describe physics beyond the SM
\cite{Pankov}. However, in our case, when we are interested in
searching for signals of the $Z'$-boson on the base of the
effective low-energy Lagrangian (\ref{1a}), these contributions to
the cross-section are essential.

\section{Many-parameter fits}

As the basic observable to fit the LEP2 experiment data on the
Bhabha process we propose the differential cross-section
\begin{equation}\label{7}
\left.\frac{d\sigma^\mathrm{Bhabha}}{dz}-\frac{d\sigma^{\mathrm{Bhabha},SM}}{dz}\right|_{z=z_i,\sqrt{s}=\sqrt{s_i}},
\end{equation}
where $i$ runs over the bins at various center-of-mass energies
$\sqrt{s}$. The final differential cross-sections measured by the
ALEPH (130-183 GeV, \cite{ALEPH}), DELPHI (189-207 GeV,
\cite{DELPHI}), L3 (183-189 GeV, \cite{L3}), and OPAL (130-207
GeV, \cite{OPAL}) collaborations are taken (299 bins).

As the observables for $e^+e^-\to\mu^+\mu^-,\tau^+\tau^-$
processes, we consider the total cross-section and the
forward-backward asymmetry
\begin{equation}\label{8}
\sigma^{\ell^+\ell^-}_T-\sigma_T^{\ell^+\ell^-,\mathrm{SM}},
\quad
\left.A^{\ell^+\ell^-}_{FB}-A_{FB}^{\ell^+\ell^-,\mathrm{SM}}\right|_{\sqrt{s}=\sqrt{s_i}},
\end{equation}
where $i$ runs over 12 center-of-mass energies $\sqrt{s}$ from 130
to 207 GeV. We consider the combined LEP2 data \cite{EWWG} for
these observables (24 data entries for each process). These data
are more precise as corresponding differential cross-sections. Our
analysis is based on the fact that the kinematics of the
$s$-channel processes is rather simple and a differential
cross-section is effectively the two-parametric function of the
scattering angle. The total cross-section and the forward-backward
asymmetry incorporate a complete information about the kinematics
of the process and therefore are an adequate alternative for the
differential cross-sections.

The data are analysed by means of the $\chi^2$ fit. Denoting the
observables (\ref{7})--(\ref{8}) by $\sigma_i$, one can construct
the $\chi^2$-function,
\begin{equation}\label{9}
\chi^2(\bar{a}, \bar{v}_e,\bar{v}_\mu,\bar{v}_\tau) =
\sum\limits_i
\left[\frac{\sigma^\mathrm{ex}_i-\sigma^\mathrm{th}_i(\bar{a},
\bar{v}_e,\bar{v}_\mu,\bar{v}_\tau)}{\delta\sigma_i}\right]^{2},
\end{equation}
where $\sigma^\mathrm{ex}$ and $\delta\sigma$ are the experimental
values and uncertainties of the observables, and
$\sigma^\mathrm{th}$ are their theoretical expressions presented
in Eqs. (\ref{4})--(\ref{5}). The sum in Eq. (\ref{9}) refers to
either the data on one specific process or the combined data on
several processes. By minimizing the $\chi^2$-function, the
maximal-likelihood estimate for the $Z'$ couplings can be derived.
The $\chi^2$-function is also used to plot the CA in the space of
$\bar{a}$, $\bar{v}_e$, $\bar{v}_\mu$, and $\bar{v}_\tau$.

For all the considered processes, the theoretic predictions
$\sigma^\mathrm{th}_i$ are linear combinations of products of two
$Z'$ couplings
\begin{eqnarray}\label{10}
\sigma^\mathrm{th}_i&=&\sum_{j=1}^{7} C_{ij}A_j,\\
 A_j&=&\{\bar{a}^2,\bar{v}_e^2,\bar{a}\bar{v}_e,\bar{v}_e\bar{v}_\mu,\bar{v}_e\bar{v}_\tau,
\bar{a}\bar{v}_\mu,\bar{a}\bar{v}_\tau\},
\nonumber
\end{eqnarray}
where $C_{ij}$ are known numbers. In what follows we use the
matrix notation $\sigma^\mathrm{th}=\sigma^\mathrm{th}_i$,
$\sigma^\mathrm{ex}=\sigma^\mathrm{ex}_i$, $C=C_{ij}$, $A=A_j$.
The uncertainties $\delta\sigma_i$ can be substituted by a
covariance matrix $D$. The diagonal elements of $D$ are
experimental errors squared,
$D_{ii}=(\delta\sigma^\mathrm{ex}_i)^2$, whereas the non-diagonal
elements are responsible for the possible correlations of
observables. The $\chi^2$-function can be rewritten as
\begin{eqnarray}\label{11}
\chi^2(A)&=&(\sigma^\mathrm{ex}-\sigma^\mathrm{th})^\mathrm{T}
D^{-1}
(\sigma^\mathrm{ex}-\sigma^\mathrm{th})
\nonumber\\
&=&(\sigma^\mathrm{ex}-CA)^\mathrm{T}
D^{-1} (\sigma^\mathrm{ex}-CA),
\end{eqnarray}
where upperscript T denotes the matrix transposition.

The $\chi^2$-function has a minimum, $\chi^2_\mathrm{min}$, at
\begin{equation}\label{12}
\hat{A}=(C^\mathrm{T}D^{-1}C)^{-1}C^\mathrm{T}D^{-1}\sigma^\mathrm{ex}
\end{equation}
corresponding to the maximum-likelihood values of $Z'$ couplings.
From Eqs. (\ref{11}), (\ref{12}) we obtain
\begin{eqnarray}\label{13}
\chi^2(A)-\chi^2_\mathrm{min}&=& (\hat{A}-A)^\mathrm{T}
\hat{D}^{-1}(\hat{A}-A),
\nonumber\\
\hat{D}&=& (C^\mathrm{T}D^{-1}C)^{-1}.
\end{eqnarray}

Usually, the experimental values $\sigma^\mathrm{ex}$ are
normal-distributed quantities with the mean values
$\sigma^\mathrm{th}$ and the covariance matrix $D$. The quantities
$\hat{A}$, being the superposition of $\sigma^\mathrm{ex}$, also
have the same distribution. It is easy to show that $\hat{A}$ has
the mean values $A$ and the covariance matrix $\hat{D}$.

The inverse matrix $\hat{D}^{-1}$ is symmetric and can be
diagonalized. The number of non-zero eigenvalues is determined by
the rank (denoted $M$) of $\hat{D}^{-1}$. The rank $M$ equals to
the number of linear-independent terms in the observables
$\sigma^\mathrm{th}$. So, the right-hand-side of Eq. (\ref{13}) is
a quantity distributed as $\chi^2$ with $M$ degrees of freedom
(d.o.f.). Since this random value is independent of $A$, the CA in
the parameter space ($\bar{a}$, $\bar{v}_e$, $\bar{v}_\mu$,
$\bar{v}_\tau$) corresponding to the probability $\beta$ can be
defined as \cite{SMEP}:
\begin{equation}\label{14}
\chi^2\le \chi^2_\mathrm{min}+\chi^2_{\mathrm{CL},\beta}(M),
\end{equation}
where $\chi^2_{\mathrm{CL,\beta}}(M)$ is the $\beta$-level of the
$\chi^2$-distribution with $M$ d.o.f.

In the Bhabha process, the $Z'$ effects are determined by 3
linear-independent contributions coming from $\bar{a}^2$,
$\bar{v}_e^2$, and $\bar{a}\bar{v}_e$ ($M=3$). As for the
$e^+e^-\to\mu^+\mu^-,\tau^+\tau^-$ processes, the observables
depend on 4 linear-independent terms for each process:
$\bar{a}^2$, $\bar{v}_e\bar{v}_\mu$, $\bar{v}_e\bar{a}$,
$\bar{a}\bar{v}_\mu$ for $e^+e^-\to\mu^+\mu^-$; and $\bar{a}^2$,
$\bar{v}_e\bar{v}_\tau$, $\bar{v}_e\bar{a}$, $\bar{a}\bar{v}_\tau$
for $e^+e^-\to\tau^+\tau^-$ ($M=4$). Note that a number of terms
in the observables for different processes are the same.
Therefore, the number of $\chi^2$ d.o.f. in the combined fits is
less than the sum of d.o.f. for separate processes. Hence, the
predictive power of the larger set of data is not drastically
spoiled by increased number of d.o.f. In fact, combining the data
of the Bhabha and $e^+e^-\to\mu^+\mu^-$ ($\tau^+\tau^-$) processes
together we have to treat 5 linear-independent terms. The complete
data set for all the lepton processes is ruled by 7 d.o.f. As a
consequence, the combination of the data for all the lepton
processes is possible.

The parametric space of couplings ($\bar{a}$, $\bar{v}_e$,
$\bar{v}_\mu$, $\bar{v}_\tau$) is four-dimensional. However, for
the Bhabha process it is reduced to the plane ($\bar{a}$,
$\bar{v}_e$), and to the three-dimensional volumes ($\bar{a}$,
$\bar{v}_e$, $\bar{v}_\mu$), ($\bar{a}$, $\bar{v}_e$,
$\bar{v}_\tau$) for the $e^+e^-\to\mu^+\mu^-$ and
$e^+e^-\to\tau^+\tau^-$ processes, correspondingly. The predictive
power of data is distributed not uniformly over the parameters.
The parameters $\bar{a}$ and $\bar{v}_e$ are present in all the
considered processes and appear to be significantly constrained.
The couplings $\bar{v}_\mu$ or $\bar{v}_\tau$ enter when the
processes $e^+e^-\to\mu^+\mu^-$ or $e^+e^-\to\tau^+\tau^-$ are
accounted for. So, in these processes, we also study the
projection of the CA onto the plane ($\bar{a},\bar{v}_e$).

The beginning of the parametric space, $\bar{a}=\bar{v}_e=0$,
corresponds to the absence of the $Z'$ signal. This is the SM
value of the observables. This point could occur inside or outside
of the CA at a fixed CL. When it lays out of the CA, this means
the distinct signal of the Abelian $Z'$. Then the signal
probability can be defined as the probability that the data agree
with the Abelian $Z'$ boson existence and exclude the SM value.
This probability corresponds to the most stringent CL (the largest
$\chi^2_\mathrm{CL}$) at which the point $\bar{a}=\bar{v}_e=0$ is
excluded. If the SM value is inside the CA, the $Z'$ boson is
indistinguishable from the SM. In this case, upper bounds on the
$Z'$ couplings can be determined.

\begin{figure}
\centering
  \includegraphics[bb= 0 0 540 300, width=.4\textwidth]{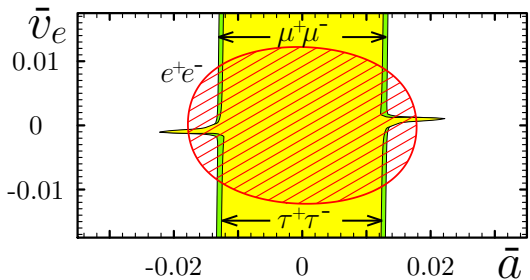}
  \caption{The 95\% CL areas in the ($\bar{a},\bar{v}_e$) plane for the Bhabha,
  $e^+e^-\to\mu^+\mu^-$, and $e^+e^-\to\tau^+\tau^-$ processes.}\label{fig:1}
\end{figure}

The 95\% CL areas in the ($\bar{a},\bar{v}_e$) plane for the
separate processes are plotted in Fig. \ref{fig:1}. As it is seen,
the Bhabha process constrains both the axial-vector and vector
couplings. As for the $e^+e^-\to\mu^+\mu^-$ and
$e^+e^-\to\tau^+\tau^-$ processes, the axial-vector coupling is
significantly constrained, only. The CAs include the SM point at
the meaningful CLs, so the experiment could not pick out clearly
the Abelian $Z'$ signal from the SM. An important conclusion from
these plots is that the experiment significantly constrains only
the couplings entering sign-definite terms in the cross-sections.

\begin{figure}
\centering
  \includegraphics[bb= 0 0 540 470 ,width=.4\textwidth]{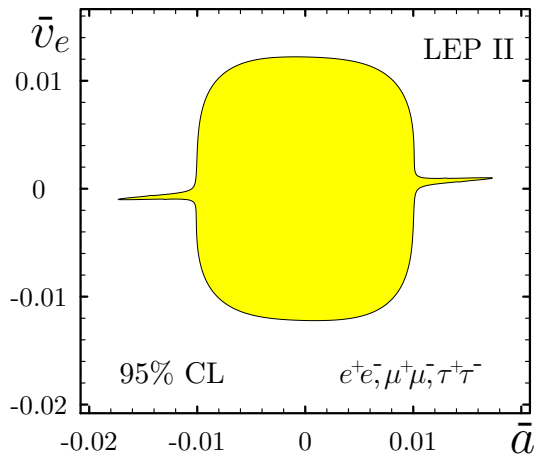}\\
  \caption{The projection of the 95\% CL area onto the ($\bar{a},\bar{v}_e$) plane
  for the combination of the Bhabha, $e^+e^-\to\mu^+\mu^-$, and $e^+e^-\to\tau^+\tau^-$
  processes.}\label{fig:2}
\end{figure}

The combination of all the lepton processes is presented in Fig.
\ref{fig:2}. There is no visible signal beyond the SM. The
couplings to the vector and axial-vector electron currents are
constrained by the many-parameter fit as $|\bar{v}_e|<0.012$,
$|\bar{a}|<0.018$ at the 95\% CL. If the charge corresponding to
the $Z'$ interactions is assumed to be of order of the
electromagnetic one, then the $Z'$ mass should be greater than
0.67 TeV. For the charge of order of the SM $SU(2)_L$ coupling
constant $m_{Z'}\ge 1.4$ TeV. One can see that the constraint is
not too severe to exclude the $Z'$ searches at the LHC.

Let us compare the obtained results with the one-parameter fits in
Ref. \cite{PRD2004}. Fitting the current data with the
one-parameter observable, we find the updated values of the $Z'$
coupling to the electron vector current together with their
1$\sigma$ uncertainties:
\begin{eqnarray}
  \mathrm{ALEPH}: &\bar{v}_e^2=& -0.11\pm 6.53 \times 10^{-4}\nonumber\\
  \mathrm{DELPHI}: &\bar{v}_e^2=& 1.60\pm 1.46 \times 10^{-4} \nonumber\\
  \mathrm{L3}: &\bar{v}_e^2=& 5.42\pm 3.72 \times 10^{-4} \nonumber\\
  \mathrm{OPAL}: &\bar{v}_e^2=& 2.42\pm 1.27 \times 10^{-4} \nonumber\\
  \mathrm{Combined}: &\bar{v}_e^2=& 2.24\pm 0.92 \times 10^{-4}. \nonumber
\end{eqnarray}
As it is seen, the most precise data of DELPHI and OPAL
collaborations are resulted in the Abelian $Z'$ signals at one and
two standard deviation level, correspondingly. The combined value
shows the 2$\sigma$ signal, which corresponds to $0.006\le
|\bar{v}_e|\le 0.020$.

\begin{figure}
\centering
  \includegraphics[bb= 0 0 540 390,width=.4\textwidth]{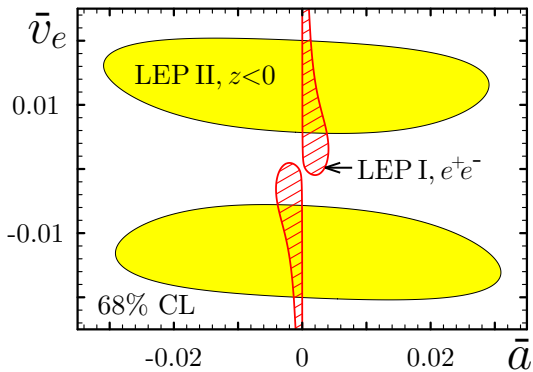}\\
  \caption{The 68\% CL area in the ($\bar{a},\bar{v}_e$) plane
  from the backward bins of the Bhabha process in the LEP2 experiments
  (the shaded area). The hatched area is the 68\% CL area from
  the LEP 1 data on the Bhabha process.}\label{fig:3}
\end{figure}

On the other hand, our many-parameter fit constrains the $Z'$
coupling to the electron vector current as $|\bar{v}_e|\le 0.012$
with no evident signal. Why does the one-parameter fit of the
Bhabha process show the 2$\sigma$ CL signal whereas there is no
signal in the two-parameter one? Our one-parameter observable
accounts mainly for the backward bins. This is in accordance with
the kinematic features of the process. Therefore, the difference
of the results can be inspired by the data sets used. To check
this, we perform the many-parameter fit with the 113 backward bins
($z\le 0$), only. The $\chi^2$ minimum,
$\chi^2_\mathrm{min}=103.0$, is found in the non-zero point
$|\bar{a}|=0.00017$, $\bar{v}_e= 0.015$. This value of the $Z'$
coupling $\bar{v}_e$ is in an excellent agreement with the mean
value obtained in the one-parameter fit. The 68\% CA in the
($\bar{a},\bar{v}_e$) plane is plotted in Fig. \ref{fig:3}. There
is a visible signal of the Abelian $Z'$ boson. The zero point
$\bar{a}=\bar{v}_e=0$ (the absence of the $Z'$ boson) corresponds
to $\chi^2=107.9$. It is covered by the CA with $1.33\sigma$ CL.
Thus, the backward bins show the $1.33\sigma$ signal of the
Abelian $Z'$ boson in the many-parameter fit. So, the
many-parameter fit is less precise than the analysis of the
one-parameter observables.

At LEP1 experiments \cite{LEP1} the $Z$-boson coupling constants
to the vector and axial-vector lepton currents ($g_V$, $g_A$) were
precisely measured. The Bhabha process shows the 1$\sigma$
deviation from the SM values for Higgs boson masses $m_H\ge 114$
GeV (see Fig. 7.3 of Ref. \cite{LEP1}). This deviation could be
considered as the effect of the $Z$--$Z'$ mixing. It is
interesting to estimate the bounds on the $Z'$ couplings following
from these experiments.

Due to the RG relations, the $Z$--$Z'$ mixing angle is completely
determined by the axial-vector coupling $\bar{a}$. So, the
deviations of $g_V$, $g_A$ from their SM values are governed by
the couplings $\bar{a}$ and $\bar{v}_e$,
\begin{equation}\label{lep1}
g_V-g_V^{\mathrm{SM}}=-49.06 \bar{a}\bar{v}_e,\quad
g_A-g_A^{\mathrm{SM}} = 49.06 \bar{a}^2.
\end{equation}
Let us assume that a total deviation of theory from experiments
follows due to the $Z$--$Z'$ mixing. This gives an upper bound on
the $Z'$ couplings. In this way one can estimate is the $Z'$ boson
excluded by the experiments or not.

The 1$\sigma$ CL area for the Bhabha process from Ref. \cite{LEP1}
is converted to the ($\bar{a},\bar{v}_e$) plane in Fig.
\ref{fig:3}. The SM values of the couplings correspond to the top
quark mass $m_t=178$ GeV and the Higgs scalar mass $m_H=114$ GeV.
As it is seen, the LEP1 data on the Bhabha process is compatible
with the Abelian $Z'$ existence at the $1\sigma$ CL. The
axial-vector coupling is constrained as $|\bar{a}|\le 0.005$. This
bound corresponds to $\bar{a}^2\le 2.5\times 10^{-5}$, which
agrees with our previous one-parameter fits of the LEP2 data for
$e^+e^-\to\mu^+\mu^-,\tau^+\tau^-$ processes \cite{YAF2004}
($\bar{a}^2= 1.3\pm 3.89\times 10^{-5}$ at 68\% CL). On the other
hand, the vector coupling constant $\bar{v}_e$ is practically
unconstrained by the LEP1 experiments.

\section{DISCUSSION}

LEP collaborations have reported about a good agreement between
the experimental data and the predictions of the SM
\cite{ALEPH,DELPHI,L3,OPAL}. Our analysis of the leptonic
processes based on the same data set and the same SM values of
cross-sections showed that the existence of $Z'$ boson with the
mass of order 1-1.2 TeV is not excluded at the 1-2$\sigma$ CL. We
observed this in one-parameter fits \cite{YAF2004,PRD2004} and in
the many-parameter fits in the present paper. The estimated $Z'$
parameters derived in different methods are in good agreement with
each other. So, we are faced with a necessity to find a possible
explanations of this discrepancy. We believe that the reason is in
the RG relations, which play a crucial role in treating
experimental data. As we showed, the RG relations served to reduce
the number of unknown parameters and extract a maximal information
about the signal of the particle of interest from the experimental
data set. If the RG relations are not taken into account, no
signals of $Z'$ will be found.

LEP collaborations performed also model-dependent fits concerning
popular $Z'$ models. Since these models are included in the ones
suiting the RG relations (\ref{3}), it is interesting to compare
their analysis with our results. In experiments
\cite{EWWG,ALEPH,DELPHI,L3,OPAL} the low bound on the $Z'$ mass
was obtained. It varies from 400 to 800 GeV at the 95\% CL
dependently on the $Z'$ model. These bounds allow the $Z'$ boson
with the mass of order 1 TeV, being compatible with our results. A
possibility to select $Z'$ signals in specific scattering
processes was not discussed in these papers. This is the reason
why no signals were observed.

In our analysis we treat the data for leptonic processes only.
LEP2 collaborations measured also the total cross-sections of the
electron-positron annihilation into quark-antiquark pairs. The
Abelian $Z'$ signal in $e^+e^-\to\bar{q}q$ process is
characterized by 5 independent parameters (for example, two $Z'$
couplings to electron, $\bar{a}$ and $\bar{v}_e$, and three $Z'$
couplings to $d$, $s$ and $b$ quarks). There are 8 linear
independent terms in the cross-section. As one can check, the
experimental statistics of 12 cross-sections for different
center-of-mass energies is completely insufficient to constrain
significantly the parameters of the $Z'$ boson.

As we have shown in Ref. \cite{PRD2004}, there is the $2\sigma$
signal of the Abelian $Z'$ boson in the one-parameter fit of LEP2
data for the Bhabha process. This result is reproduced in the
present paper by fitting the updated experimental data. In the
present analysis we applied many-parameter fits for leptonic
processes for different sets of bins included. In particular, for
the backward bins (responsible for the signal due to the
kinematics of the process) the $1.3\sigma$ signal of the particle
is found. The fit of the complete set of bins constrains the $Z'$
couplings to vector and axial-vector electron currents allowing
the $Z'$ boson with the mass of order 1 TeV. Thus, we have to
conclude that the LEP2 data allow the existence of quite light
$Z'$-boson, which has a chance to be discovered in the nearest
future, at the LHC. We believe that the RG relations used in the
present analysis will be also important in searches for the $Z'$
boson at the LHC.

\section*{ACKNOWLEDGEMENT}

The authors are grateful to Pat Ward from the OPAL Collaboration
for the detailed information about the experimental data.

One of the authors (AG) thanks the ICTP (Trieste, Italy) for kind
hospitality when the paper was prepared.

This work is supported by the grant F7/296-2001 of the Fundamental
Researches State Fund of Ukraine.

\end{document}